\documentclass[usenatbib,usegraphicx,useAMS]{mn2e}
\usepackage{times}

\title[Effects of asphericity on weak lensing masses]{Effects of asphericity and substructure on 
the determination of cluster mass with weak gravitational lensing}

\author[D.~Clowe, G.~De Lucia, L.~King]{D.~Clowe,$^1$\thanks{Currently at Steward Observatory, University of Arizona} G.~De Lucia,$^2$  L. King$^1$ \\
$^1$Institut f\"ur Astrophysik und Extraterrestrische Forschung der 
Universit\"at Bonn, Auf dem H\"ugel 71, 53121 Bonn, Germany \\
$^2$Max-Planck-Institut f\"ur Astrophysik, Karl Schwarzschild Str. 1, 85741 
Garching, Germany}

\begin{document}

\date{Accepted 3000 January 1. Received 3000 January 1}

\pagerange{\pageref{firstpage}--\pageref{lastpage}} \pubyear{2003}

\maketitle

\label{firstpage}

\begin{abstract}
Weak gravitational lensing can be used to directly measure the mass along a
line-of-sight without any dependence on the dynamical state of the mass, and
thus can be used to measure the masses of clusters even if they are not
relaxed.  One common technique used to measure cluster masses 
is fitting azimuthally-averaged gravitational shear profiles with a spherical 
mass model.  In
this paper we quantify how asphericity and projected substructure in clusters
can affect the virial mass and concentration measured with this
technique by simulating weak lensing observations on 30 independent
lines-of-sights through each of four high-resolution N-body cluster 
simulations.  We find that the variations in the measured 
virial mass and concentration are of a size similar to the error expected
in ideal weak lensing observations and are correlated, but that the virial
mass and concentration of the mean shear profile agree well with that measured
in three dimensional models of the clusters.  The dominant effect
causing the variations is the proximity of the line-of-sight to the major
axis of the 3-D cluster mass distribution, with projected substructure only
causing minor perturbations in the measured concentration.  Finally we find
that the best-fit ``universal'' CDM models used to fit the shear profiles 
over-predict the surface density of the clusters due to the
cluster mass density falling off faster than the $r^{-3}$ model assumption.
\end{abstract}


\section{Introduction}

Weak gravitational lensing, in which mass in a field is measured by the 
distortion induced in the shapes of background galaxies, has proven to be
a powerful tool in the study of clusters \citep[see reviews by][]{BA01.1,ME99.1}.
With the advent of wide-field, multi-chip CCD cameras,
weak lensing shear profiles for clusters have been measured to beyond the
virial radius \citep{CL01.1,CL02.1,DA02.1} with a high signal-to-noise.  
For many of these clusters, however, there is a disagreement in the cluster
mass as measured by weak lensing and by strong lensing, X-ray observations,
and cluster galaxy velocity dispersions.

One possible origin for the differences in the mass estimates is the error
introduced by fitting spherically symmetric mass models to aspherical
structures.  \citet{PI03.1} have calculated that imposing a spherical model
on a smooth tri-axial cluster can change the ratio of the measured X-ray 
mass to weak lensing mass by up to $30\%$.  \citet{KI01.1} investigated
the effect of small-scale substructure seen in N-body simulations of clusters 
and concluded that the departures from a smooth mass model caused by these 
substructures do not greatly effect the weak lensing mass measurements.

Another possible origin for the mass estimate differences is the projection
of mass structures along the line-of-sight onto the cluster mass in the weak
lensing measurements.  Foreground and background structures, for which there
exists no positional correlation with the cluster, do not produce a bias
in the weak lensing measurements \citep{HO03.1}.  Filamentary structures
extending from the cluster along the line-of-sight can potentially cause 
an overestimate of the cluster mass from weak lensing, with estimates of the
additional mass measured in N-body simulations ranging from a few percent 
\citep{CE97.1, RE99.1} to over $50\%$ \citep{ME01.1}.  However, these results
are obtained by comparing the total mass projected in a cylinder to that
contained in a sphere in the N-body simulation, and not by fitting the shear
produced by the projected mass with a projected mass model, as is most
commonly done for weak lensing mass determinations of clusters.  

In this paper we use four high-resolution N-body simulations of massive 
clusters to study the effects of cluster asphericity, secondary halos, and 
filamentary structures on the mass profiles measured by weak lensing.
In Section 2 we present the simulations and methods used
to project the 3-dimensional simulations to 2-dimensional mass maps.  We
discuss the weak lensing techniques and results in Section 3, and present
our conclusions in Section 4.  Throughout this paper we assume the cosmology
of the simulations ($\Omega_\mathrm{m} = 0.3$, $\Omega_{\Lambda} = 0.7$,
$H_0 = 70$ km/s/Mpc, spectral shape $\Gamma=0.21$, 
and spectral normalization $\sigma_8=0.9$).

\section{Simulations}

The simulations used in this work were carried out by Barbara Lanzoni
as part of her PhD thesis and is described in \citet{LA03.2} and 
\citet{DE03.1}.  A suitable target cluster is selected from a 
previously generated cosmological simulation of a large region. 
The particles in the final cluster and their 
closest surroundings are traced back to their Lagrangian region; the original 
particles are replaced with a larger number of lower mass particles and 
perturbed using the same fluctuation spectrum of the parent simulation, but 
now extended to smaller scales (because of the increased dynamical range). 
Outside this high resolution region, particles with increasing mass are 
used in order to model the large-scale density and velocity field of the parent
simulation.
Assuming these new initial conditions, the particle evolution is followed using
the code \emph{GADGET}; a full description of the numerical and algorithmic 
details is given in \citet{SP01.1}.

\begin{figure*}
\centering
\includegraphics[width=14cm]{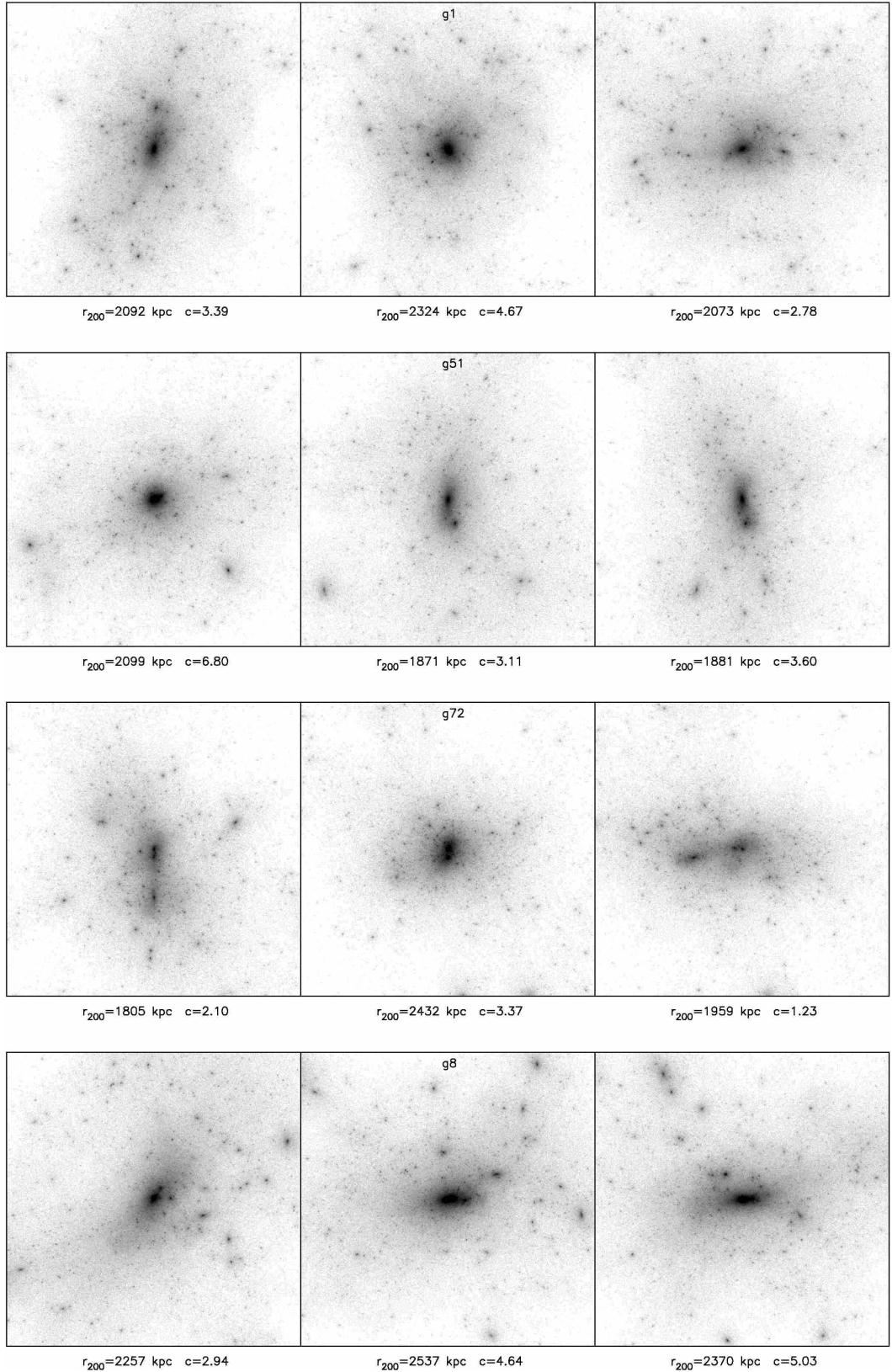}
\caption{Above are images of the surface mass density for the four 
simulations.  Each simulation is shown with three orthogonal projections,
with the left-hand panel projection being rotated about the horizontal axis 
to obtain the middle panel projection, which is then rotated about the vertical
axis to obtain the right-hand panel projection (the horizontal axis of the
left-hand image is the vertical axis in the right-hand image).  All of the
images are shown in a $\sqrt{\log}$ stretch to display structures in both
the core and near the image edge.}
\label{fig1}
\end{figure*}

In this work we use four high resolution re-simulations of clusters 
(named g1, g51, g72, and g8) with masses $\simeq 1$-$2\times 10^{15} 
M_{\odot}$. The parent simulation employed is the VLS simulation carried
out by the Virgo Consortium \citep{JE01.1,YO01.1}.  The simulation was
performed using a parallel P3M code \citep{MA98.1} and followed $512^3$ particles
with a particle mass of $7\times 10^{10}\,h^{-1}\,{\rm M}_{\odot}$ in a
comoving box of size $479\,h^{-1}$Mpc on a side.  We selected
an output of the resimulations with total elapsed time equivalent to
placing the clusters at $z_{\mathrm{cl}} = 0.2$.  This redshift was 
chosen to match current weak lensing observations of clusters with wide-field cameras.
The particle mass in the high resolution region of the final re-simulation 
is $\simeq 2\times10^9 h^{-1}M_{\odot}$.  
A standard friends-of-friends algorithm with a linking length $b=0.2$ was used
to find the group of virialized particles.  It has been shown that this linking length 
results in the selection of groups whose overdensity is close to the one predicted by the 
spherical collapse model \citep{CO96.1}.

For each simulation, 30 surface density maps were created by taking 10
independent rotations of the simulations and three orthogonal projections
for each rotation and projecting all of the particles
in a 13 Mpc cubic box, whose center was the most bound particle of the
primary halo.  The rotation of the simulations was performed before the
selection of particles inside the box.  The 13 Mpc box was the maximum 
size which was fully populated by the
high-resolution particles in all of the rotations.
The positions of the particles were projected onto a grid of 
$1300\times1300$ pixels (each pixel corresponds to 10 kpc) and a triangular shaped 
cloud \citep{HO88.1} weighting function was used to assign the mass to each 
pixel.  The nearest grid point is assigned a weight of $0.75-dx^2$ and surrounding 
points are assigned a weight of $0.5*(1.5-dx)^2$, where $dx$ is the distance 
from the sample to the grid point in units of the pixel size.

Shown in Fig.~\ref{fig1} are three orthogonal projections for each 
cluster, chosen so that
the three projections have a large difference in the resulting best-fit
surface density profiles.  All of the clusters are best described with a
triaxial mass model in 3-D, and the direction of the major axis of the mass 
distribution is fairly constant at all radii with the exception of g72, in
which the mass distribution is best described as a combination of two triaxial
systems with major axes closely aligned but offset from one another.
The major axes were calculated from the eigenvectors of the matrix $M_{\alpha \beta}$:
\begin{equation}
M_{\alpha \beta} = \sum_i=1^N X_{\alpha}^i X_{\beta}^i
\end{equation}
with $\alpha{\rm,}\beta = 1,2,3$ and $X_{\alpha}^i$ being the coordinate of the
$i$th particle with respect to the $\alpha$ axis, relative to the center of
mass.  For simplicity we fixed the center of mass on the most bound
particle of each halo and considered all the particles inside the virial
radius.

\section{Weak Lensing Analysis}

The goal of weak lensing observations is to measure the dimensionless
surface mass density of the clusters, $\kappa$, where
\begin{equation}
\kappa = {\Sigma \over \Sigma _{\mathrm{crit}}}.
\end{equation}
$\Sigma $ is the two-dimensional surface density of the cluster, and 
$\Sigma _{\mathrm{crit}}$ is a scaling factor:
\begin{equation}
\Sigma _{\mathrm{crit}} = {c^2 \over 4 \pi  G}{D_{\mathrm{s}} \over 
D_{\mathrm{l}} D_{\mathrm{ls}}}
\end{equation}
where $D_{\mathrm{s}}$ is the angular distance to the source (background) 
galaxy, $D_{\mathrm{l}}$ is the angular distance to the lens (cluster), 
and $D_{\mathrm{ls}}$ is the angular
distance from the lens to the source galaxy.  

The surface density $\kappa$ cannot, however, be measured directly from the
shapes of the background galaxies.  Instead, one can measure the mean
distortion of the galaxies by looking for a systematic deviation from
a zero average ellipticity.  From the distortion one can measure the
reduced shear $g$, which is related to the shear $\gamma$ by
\begin{equation}
g = {\gamma \over 1-\kappa}.
\label{eq3}
\end {equation}
Once the reduced shear is measured from the background galaxy ellipticities,
one can then convert the shear measurements to $\kappa $ measurements, and
then to surface mass measurements if one knows the redshifts of the lens
and background galaxies, using a variety of techniques \citep{BA01.1,ME99.1}.

The technique which we are testing with simulations is that of parameterized
model fitting, in which the azimuthally-averaged measured shear is fit with radial 
surface mass profiles from chosen model families.  The radial mass profiles
are first converted to $\kappa $ profiles by assuming a mean redshift for the
background galaxies, and then to a reduced shear profile using
\begin{equation}
\langle \gamma (r)\rangle = \bar{\kappa}(r) - \langle \kappa (r)\rangle
\label{eq4}
\end{equation}
where $\langle \rangle$ indicate the azimuthally averaged quantities and
$\bar{\kappa}(r)$ is the mean $\kappa $ within radius $r$.  From 
Eqns.~\ref{eq3} and \ref{eq4},
one then has
\begin{equation}
\langle g(r)\rangle = {\bar{\kappa}(r) - \langle \kappa (r)\rangle \over
1 - \langle \kappa (r)\rangle }.
\label{eq5}
\end{equation}
This is strictly true only for a circular surface mass profile, but should
be a good approximation if the change in $\kappa (r)$ along the 
averaging circle is small compared to 1.  The model reduced shear profile
is then compared to the measured profile and the parameters of the model varied
to obtain the best fit.

\begin{figure}
\centering
\resizebox{\hsize}{!}{\includegraphics{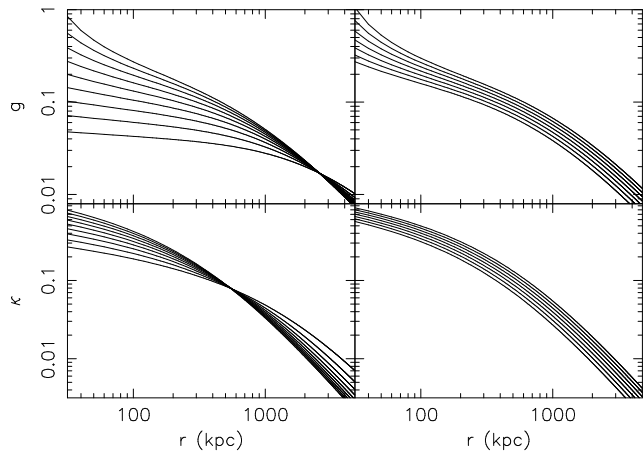}}
\caption{Shown above are the reduced shear, $g$, and convergence, $\kappa $,
profiles for NFW models in which one of the two parameters is varied.  In
the left-hand panels, $c$ varies from $1.0$ to $5.0$ in increments of $0.5$,
with $r_{200}$ kept fixed at 2000 kpc.
In the right-hand panels, $r_{200}$ varies from 1800 kpc to 2400 kpc in 
increments of $100$ kpc, with $c$ kept fixed at 4.0.  As can be seen,
increasing $r_{200}$ causes a general increase of both $\kappa$ and $g$ at
all radii, while increasing $c$ causes an increase in both $\kappa$ and $g$
at small radii, but a decrease at large radii.}
\label{fig2}
\end{figure}

The model family we fit to the data is the ``universal CDM profile''
from \citet[hereafter NFW]{NA97.6}.  These profiles have a density profile
given by
\begin{equation}
\rho (x) = {\delta _c \rho _\mathrm{c} \over x(1+x)^2}
\label{eq6}
\end{equation}
where $x = r c/r_{200}$ is a dimensionless radius based on the collapse radius,
$r_{200}$ (defined as the radius inside which the mass density is equal to
200 times the critical density, $\rho _\mathrm{c}$) and the concentration, 
$c$, and $\delta _c$ is a scaling factor which depends on $c$.
Formulas for the surface density, obtained by integrating Eq.~\ref{eq6} 
along the line of sight, and resulting shear profile can be 
found in \citet{BA96.1} and \citet{WR00.1}, and the
reduced shear profile calculated using Eq.~\ref{eq3}.

Shown in Fig.~\ref{fig2} are the effects of changing $r_{200}$ and $c$ on the reduced
shear profiles and $\kappa $ profiles of a cluster.  Increasing $r_{200}$
results in an increase of both $\kappa$ and $g$ at all radii.  Increasing $c$,
however, results in a shift of mass from the outer parts of the cluster into
the core, steepening the rate at which $\kappa $ decreases with radius.
Because the reduced shear measures the change in surface density, 
increasing the value of $c$ greatly increases the shear of the cluster at
all radii inside $r_{200}$.  If two spherical NFW mass structures
are superimposed by projection, the measured $r_{200}$ will be relatively insensitive
to the alignment of the cores.  The measured concentration, however, will depend 
strongly on how well aligned the cores of the two structures are along the 
line of sight, with structures which are misaligned by a substantial fraction of
the combined $r_{200}$ being detected at extremely low concentration.

\begin{figure}
\centering
\resizebox{\hsize}{!}{\includegraphics{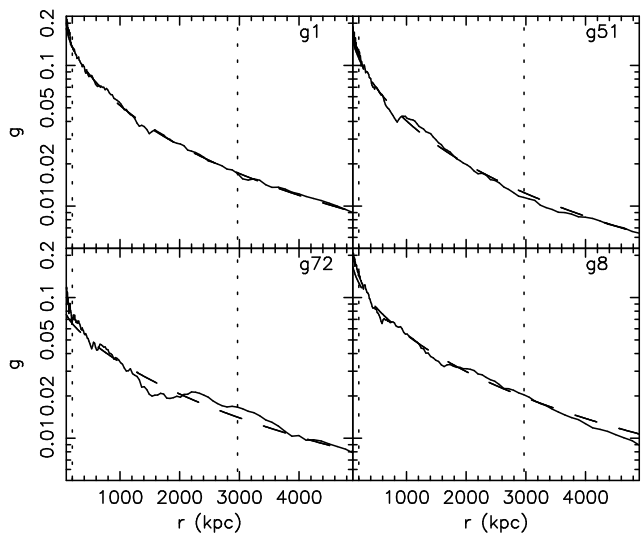}}
\caption{Shown above are the reduced shear profiles for one projection from
each of the four simulations.  The solid line is the reduced shear from the
projection while the dashed line is the best fitting NFW profile.  The dotted
lines indicate the region over which the fit was performed.}
\label{fig3}
\end{figure}

\subsection{Best-fit NFW profiles}
In order to measure the best-fit NFW profiles, the 30 surface density 
projections for each cluster were converted to $\kappa $-maps assuming the
clusters are at $z_\mathrm{cl} = 0.2$ and the background galaxies lie on a
sheet at $z_\mathrm{bg} = 1.0$.  The $\kappa $-maps were then converted to
shear maps by utilizing the fact that both are combinations of second 
derivatives of the surface potential, and therefore
\begin{equation}
\tilde{\gamma } = \left( {\hat{k}_1^2 - \hat{k}_2^2 \over \hat{k}_1^2 +
\hat{k}_2^2} \tilde{\kappa}, {2 \hat{k}_1 \hat{k}_2 \over \hat{k}_1^2 +
\hat{k}_2^2} \tilde{\kappa} \right)
\end{equation}
where $\tilde{\gamma}$ and $\tilde{\kappa}$ are the Fourier transforms of the
shear and convergence, and $\hat{k}$ are the appropriate wave vectors.
The resulting shear maps are then divided by 1 minus the $\kappa$-map to
produce a reduced shear map.  These can then be discretely sampled and have 
noise added in order to simulate a background galaxy ellipticity catalog, 
and then azimuthally averaged to produce a reduced shear profile.

The two primary sources of random noise in weak lensing shear observations
are the intrinsic ellipticity distribution of the background galaxies and the
superposition of unrelated mass peaks and voids along the line of sight.  
Neither of these effects give rise to a bias in fitted models \citep{KI01.2,HO03.1}, 
and so while the best-fit model for any given noise realization can differ
significantly from the best-fit model for the noise-free shear profile, the
noise-free model is recovered when averaging over a large number of noise
realizations. As a result, we fit the reduced shear profiles for the 
simulations without adding any noise.  

\begin{figure*}
\includegraphics[width=12cm]{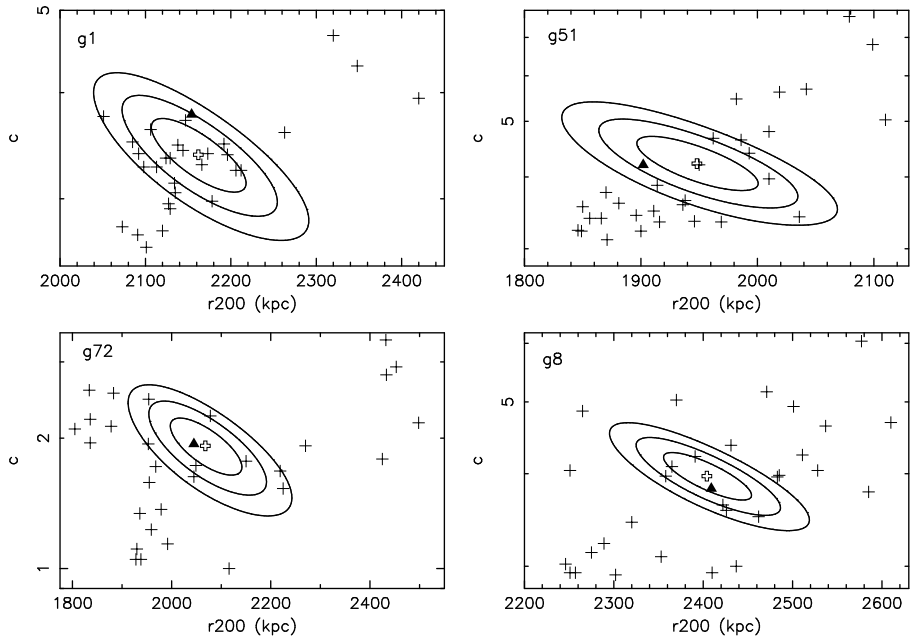}
\caption{Plotted above, as $+$, are the best fit values for $r_{200}$ and 
$c$ for the 30 projections for each simulation.  Plotted as the solid triangles
are the parameters for the 3-D fit for each simulation, and the open crosses
are the fit parameters to the mean of the reduced shear profiles for each
simulation.  The solid contours are what the 1-, 2-, and 3-$\sigma$ error 
contours for a single realization (the open cross) under 
``ideal'' observing conditions with current telescopes.}
\label{fig4}
\end{figure*}

The fits were performed by taking the 2-D shear maps and azimuthally averaging
the tangential components of the shear in logarithmically spaced radial 
annuli about the chosen center.  The radial shear profile was then fit with
projected NFW shear profiles using $\chi ^2$ minimization.
While the tangential shear did not have
any noise added, we did create a noise estimate for each annulus in order
to get a $\chi ^2$ statistic from the fitting which could be compared 
among the different projections and simulations.  The noise level was calculated
to mimic the average noise from an image which provided 100 galaxies per
square arcminute [roughly the usable number density of galaxies in the
Hubble Deep Fields \citep{ME01.2}] and 1-D rms shear noise from intrinsic
ellipticities of galaxies of $0.2$ ($0.3$ is commonly measured using second
moments, \citet{MA03.1} suggest this can be reduced to $\sim 0.2$ by including
higher order moments).  These noise estimates are therefore estimates of what
is expected from a deep, wide-field image from a space-based telescope of a
cluster which has minimal fore- and background structures superimposed.

The shear profiles were created using the location of the most bound particle,
which is assumed to be the location of the brightest cluster galaxy.  The
peak in the projected surface density was typically located within 30 kpc
of the most bound particle, and switching the center of the shear profile to
the peak position did not greatly effect the parameters of the best fit models.
The reduced shear profile was fit over the radial range of $1\arcmin$ 
(197.7 kpc) to $15\arcmin$ (2.965 Mpc), which is the typical fitting range
in recent observations with wide-field cameras \citep{CL02.1,CL01.1,DA02.1}.
Inside of $1\arcmin$ the weak lensing signal is typically lost due to crowding
from cluster galaxies preventing the measurement of shapes of background 
galaxies, while the $15\arcmin$ outer radius is determined by the field of
view of the mosaic cameras.  \citet{HO03.1} suggests that the noise due to
projection of unrelated mass structures along the line of sight will increase
rapidly at radii larger than $15\arcmin$, and therefore little additional
information on the shear profile can be obtained at larger radii.

Shown in Fig.~\ref{fig3} are the reduced shear profiles and best-fit NFW reduced
shear profiles for one projection from each of the four clusters.  As can be seen, 
the NFW models provide good fits to the simulations over these
regions, and typically are also in good agreement with the reduced shear
in the projections at radii outside of the fitting region.  

\begin{figure}
\resizebox{\hsize}{!}{\includegraphics{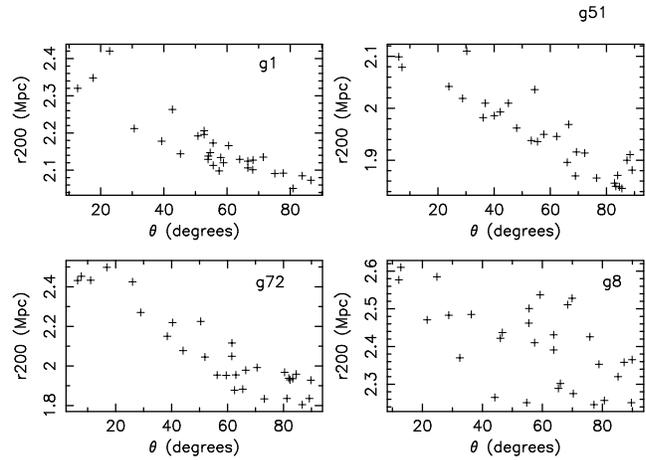}}
\caption{Plotted above are the angular offset between the line-of-sight of
the projections and the major-axis of the clusters and the best fit values for 
$r_{200}$ for the 30 projections for each simulation.}
\label{fig5}
\end{figure}

The best-fit NFW model parameters for the projections are shown in Fig.~\ref{fig4},
along with the best-fit NFW model to the 3-D density profile and the
best-fit NFW model when the shear profiles for the 30 projections are
averaged.  The parameters for the individual projections span a range of
$\pm 10$-$15\%$ in $r_{200}$ and up to a factor of 2 in $c$, and are, in
general, correlated, with higher $r_{200}$ fits also having higher $c$.
The significances of the correlation, as measured by the linear-correlation coefficient,
are $99.98\%, 99.99996\%, 95.12\%,$ and $99.79\%$ for the g1, g51, g72, and
g8 simulations respectively.  

The value of $r_{200}$ in the projections is also,
as expected, strongly correlated with the angular offset between the line-of-sight
of the projection and the major-axis of the cluster mass distribution, as can 
be seen in Fig.~\ref{fig5}.  The likelihood of the correlation, again measured
by the linear-correlation coefficient, arising from noise in uncorrelated data
are $7\times 10^{-9}, 2\times 10^{-10}, 3\times 10^{-11},$ and $4\times 10^{-3}$
for the g1, g51, g72, and g8 simulations respectively.  
The results for each 
of the simulations are discussed in detail below.

{\bf g1 -} Most of the projections (27 of the 30) have best-fit models
which are clustered in a region between $2050 < r_{200} < 2260$ kpc and
$2.5<c<3.75$, while the three outlying projections have higher values of
both $r_{200}$ and $c$.  The value of $r_{200}$ in the projections is well
correlated with the proximity of the line-of-sight through the projection to
the major axis in the 3-D mass distribution.  
Only the three projections with lines-of-sight near the major
axis, however, have measured concentrations larger than the 3-D fit,
and while the fit to the average of the 30 projections has roughly the same
$r_{200}$ as the 3-D fit, its has a markedly lower value of $c$.

{\bf g51 -} This cluster has the highest correlation between $r_{200}$ and
$c$ in the projection best-fit parameters, which are also strongly correlated
with the proximity of the light-of-sight of the projection with the major
axis of the 3-D mass distribution.  The projection with the highest measured
$r_{200}$ has a line-of-sight which is 30\degr \ from the major axis, and is
due to the projection onto the core of the substructure seen in the lower-left 
hand corner of the left-hand panel for g51 in Fig.~\ref{fig1} which is normally
projected outside of $r_{200}$ and thus not included in the mass.  The 
projections with best-fit $r_{200}$ values near that of the 3-D fit all have
lower concentrations than the 3-D fit.  The best-fit profile to the average
of the projections has a similar concentration as the 3-D fit, but a markedly
higher $r_{200}$.

{\bf g72 -} The double-core of this cluster results in relatively low
concentrations, except when the two cores are projected on top of each other.
The high $r_{200}$ projections are those with lines-of-sight near the
major axis of the 3-D mass distribution (and the axis connecting the two
cores), with the large spread in $c$ being a result of how near the two
cores are projected.  The mid $r_{200}$, low $c$ projections are those in which
the line-of-sight is near the minor axis of the larger core, while
the low $r_{200}$, (relatively) high $c$ projections have lines-of-sight
near the intermediate axis.  The $r_{200}$, c anti-correlation seen in the
small $r_{200}$ values is due primarily to the minor and intermediate axes
of the larger core being rotated to the minor and intermediate axes of the
larger-radii mass distribution.  The best-fit profile for the average of the 
projections is similar in both $r_{200}$ and $c$ to the 3-D profile.

{\bf g8 -} The large scatter in the best-fit $r_{200}$ and $c$ is a result
of the two massive filaments and numerous secondary mass peaks in the
outskirts of the simulation region.  The larger $r_{200}$ values occur
when one of the filaments is projected within $r_{200}$ of the cluster.  The
higher values of $c$ occur when one of the secondary mass peaks is projected
near the cluster core.  Due to the crescent-moon shape of the filamentary
structure, projections were only able to have one of the two filaments
projected onto the cluster, and therefore the number of high $r_{200}$
projections is larger than is found in the other simulations, which have
more linear structures, but the fractional increase in surface mass in these 
projections is smaller compared to the 3-D fit than in the other simulations.
This also results in the smallest correlation between $r_{200}$ and the angular
offset between the line-of-sight and the major axis.
The best-fit profile from the mean shear of the projections is similar in
both $r_{200}$ and $c$ to the 3-D profile.

\begin{figure}
\centering
\resizebox{\hsize}{!}{\includegraphics{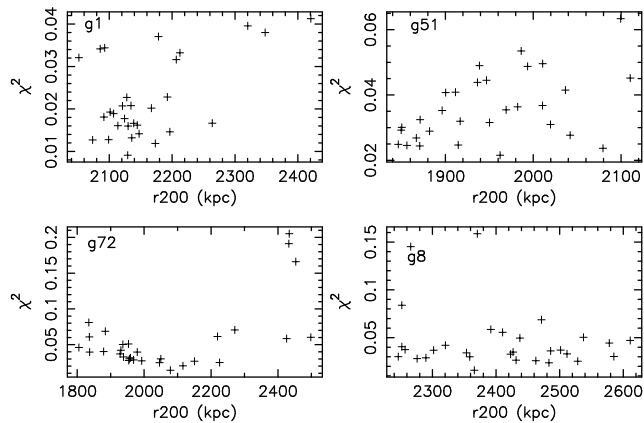}}
\caption{Plotted above are the reduced $\chi ^2$ values for the 30 projections
for each simulation.}
\label{fig6}
\end{figure}

Also shown in Fig.~\ref{fig4} are typical error contours for the NFW model fit from
the noise estimates discussed above.  As can be seen, the spread in the
best-fit parameters among the projections is slightly larger (rms of
$M_{200}$ in the projections is $\sim 30\%$ larger than the $1\sigma$
errors from the noise estimates, except for g72, for which it's almost
twice as large) than the error expected
from ``ideal'' observations which could be taken with current telescopes.
The degeneracy between $r_{200}$ and $c$ in the error contours, however, is
roughly orthogonal to the correlations from the projections, and so
the combined error contours from a single observation would tend to become 
more circular.

The quality of the fit, as measured by reduced $\chi ^2$ between the model
and projection's reduced shear profiles, is shown in 
Fig.~\ref{fig6}.  While there is a large range in the $\chi ^2$ among the
projections for a given simulation, and the g72 simulation has a higher
$\chi ^2$ on average than the rest due to the massive secondary peak,
all of the $\chi ^2$s are much smaller than 1.  Once noise is added to
the shear profile, the reduced $\chi ^2$ becomes close to 1 for all of the
projections.

\subsection{Correlation of mass with ellipticity}
One of the effects common to all four simulations is that the higher
$r_{200}$ values tend to be measured for projections in which the line-of-sight
lies near the major axis of the 3-D mass distribution.  Given that the
3-D mass distribution is generally well described by a triaxial model, one
might expect that there should exist an anti-correlation between the ellipticity 
of the central surface mass peak in the projected images and the $r_{200}$ 
measured by the fit to the shear from that peak.  The projections with the 
highest ellipticity would be those with the major axis in the plane of the 
sky, and thus have the lowest $r_{200}$.

\begin{figure}
\centering
\resizebox{\hsize}{!}{\includegraphics{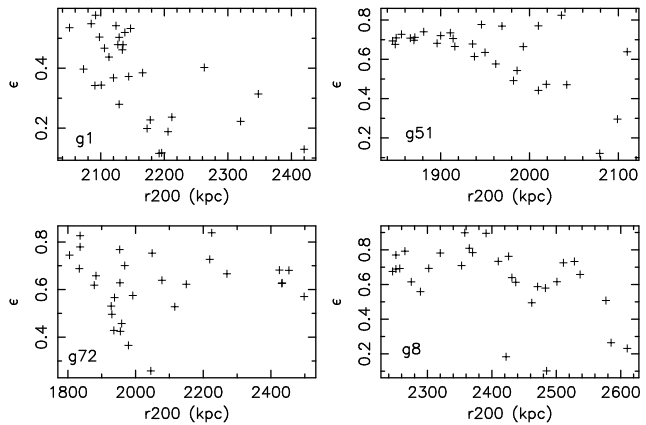}}
\caption{Plotted above are the ellipticities, as defined by Eq.~\ref{eq8} and measured
over the region $50 \mathrm{kpc}<r<500 \mathrm{kpc}$, for the 
30 projections for each simulation.  As can be seen, there is in general
an anti-correlation between the ellipticity and the measured surface mass,
but the mean ellipticity varies among the simulations.}
\label{fig7}
\end{figure}

We measured the ellipticity of the central mass peak by making a change of
variable in the
projected NFW equations from $x = r c/r_{200}$ to
\begin{equation}
\xi = x\sqrt{1-\epsilon \cos(2(\theta -\theta _0))}
\label{eq8}
\end{equation}
where $\epsilon$ gives the ellipticity and $\theta _0$ is the positional
angle.  Because this transformation of the surface mass from circularly 
symmetric to elliptical does not result in an analytic solution for the shear,
we measure the ellipticities directly in the surface density maps.
Shown in Fig.~\ref{fig7} are the ellipticities for the 30 projections plotted
against the $r_{200}$ for the best-fit NFW model.  The ellipticities
were calculated as the best fit single value of $\epsilon$ over the range of
50 to 500 kpc.  As can be seen, there is
in general an anti-correlation between ellipticity and $r_{200}$ with
higher $r_{200}$ generally resulting in lower ellipticity.  The significance
of this correlation, as measured by the linear-correlation coefficient,
is $99.6\%, 99.3\%, 74.3\%,$ and $95.0\%$ for g1, g51, g72, and g8 respectively.
The range in the
values of $\epsilon$ differ for the four clusters, however, and thus there is
not a relation which can be used as a generic correction for the observed
surface mass.

\begin{figure}
\centering
\resizebox{\hsize}{!}{\includegraphics{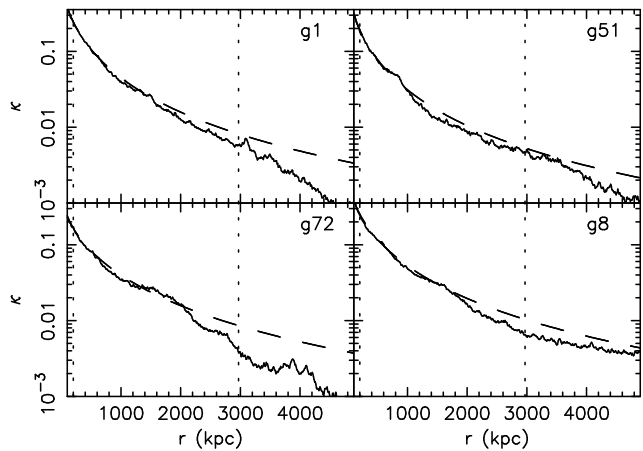}}
\caption{Shown above are the $\kappa$ profiles for one projection from
each of the four simulations, the same projections whose reduced shear
profiles are shown in Fig.~\ref{fig3}.  The solid line is the convergence from the
projection while the dashed line is the best fitting NFW profile to the
reduced shear.  The dotted lines indicate the region over which the fit was 
performed.}
\label{fig8}
\end{figure}

\begin{figure}
\centering
\resizebox{\hsize}{!}{\includegraphics{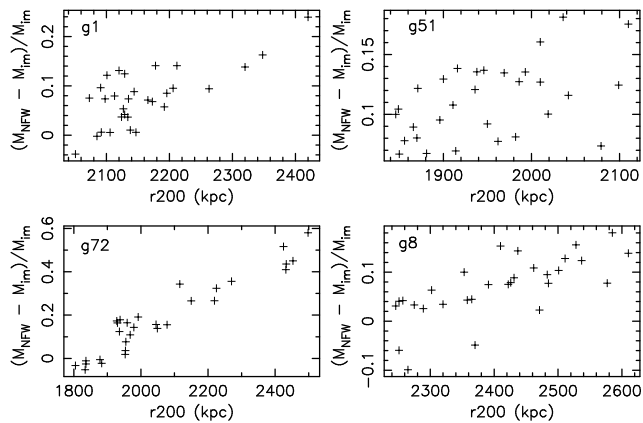}}
\caption{Plotted above are the amounts of excess surface mass within
$r\le r_{200}$ of the best-fit NFW profile to the reduced shear compared
with what is actually present in the projection, as a fraction of the
surface mass in the projection.}
\label{fig9}
\end{figure}

\subsection{Missing surface mass}
While the projected NFW profiles provide a good fit to the shear profiles
of the clusters, they predict a larger surface density than is observed at
large radii.  In Fig.~\ref{fig8} are the radial $\kappa $ profiles for the four
simulations whose shears are plotted in Fig.~\ref{fig3}, along the with 
the $\kappa $ profiles for
the NFW profiles.  As can be seen, the surface density falls off faster at 
large radius than is predicted by the best-fit NFW profile to the shear.  
This effect is seen in most of the projections for all four simulations.

In Fig.~\ref{fig9} is plotted the difference in the total surface mass within $r_{200}$
predicted by the best-fit NFW model and that actually present in the
projection.  There is a strong correlation between the excess surface mass
predicted by the NFW model and the best-fit $r_{200}$ of the model, in that
the more massive models have a greater excess surface mass.  This is a result
of the mass density decreasing faster with increasing radius along the
minor and intermediate axes than the major axis of the 3-D mass profile.
Because the high $r_{200}$ projections are those with lines-of-sight near
the 3-D major axis, the surface density in these projections has the fastest
decrease with increasing radius (which is a major cause of the increased
$\chi ^2$ in the fits, see Fig.~\ref{fig7}).  As a result, the surface density has
the greatest overestimate at large radius, and therefore these projections have
a larger overestimate of the total surface mass within $r_{200}$.  For a few
projections, this effect is mitigated by the presence of a massive secondary
peak or filamentary mass located near the $r_{200}$ outer boundary.  

The greatest discrepancy is seen in the cluster g72, which has a large
secondary core as the cluster is currently undergoing a major merger event.
This secondary core, located $1.65$ Mpc from the primary core and therefore
within the NFW fit region, results is a low value for the best-fit
concentration.  While the mass within a sphere with radius $r_{200}$ is
independent of the concentration, the total surface mass within a circle
of radius $r_{200}$ increases with decreasing concentration.  As a result,
the low concentration, caused by the presence of the second core, results in
a larger over-estimate of the surface density at large radius than is found in
the other clusters which have more typical concentration parameters.

The NFW profile has been defined in N-body simulations by only considering
the mass density profile within a sphere of radius $r_{200}$ \citep{NA97.6}.
It is therefore not surprising that the mass density profile at larger
radii might have a steeper decline than the $r^{-3}$ in the NFW profile.
Weak and strong gravitational lensing masses, however, measure the entire
mass along the line of sight, including that mass which is not gravitationally
bound to the cluster.  As such, in order to compare surface densities from
lensing to mass estimates from other means, the mass profile outside of
$r_{200}$ needs to be included in the models.  

These simulations, however, were selected to not have another massive
structure within the re-simulated region, and therefore might be biased
toward clusters surrounded by an under-dense region.  Consequently, the mass profile
at large radii might have a steeper profile for these simulations than
normal.

\section{Discussion}
While the best-fit NFW models did, on average, provide a good estimate of
the cluster virial mass and mass profile within $r_{200}$, they overestimated 
the surface mass density in the projections for radii as small as half the
virial radius.  This is due to the mass density
at large radii falling faster than the $\propto r^{-3}$ predicted by the NFW
profile, and while the mass at radii larger than the virial radius are not
considered in 3-D models, they constitute a significant fraction of the surface
density of the clusters at radii smaller than the virial radius.
The greatest discrepancy is seen in the merging cluster system g72,
which has unusually low values of the concentration, and therefore a greater
overestimate of the mass at large radius.

The cluster for which the best-fit of the mean shear field does not have
a similar $r_{200}$ to the 3-D fit is g51, in which the 2-D fit has a
$\sim 2\%$ higher value of $r_{200}$ than the 3-D fit.  This cluster has
a high ellipticity without any significant substructure at large radius
and is in virial equilibrium.  Under the tests of \citet{JI00.1}, this
cluster is one in which the NFW profile should provide a measurement of
the total mass of the system.  However, as a result of the high ellipticity, 
combined with the under-density of mass
at large radius in these simulations, the mass density decreases with 
increasing radius much more rapidly along the minor and intermediate axes of 
the cluster than is assumed by the NFW model.  As such, the sphere enclosed
by the 3-D $r_{200}$ includes a large volume with densities far below those
predicted by the NFW model.  Therefore, the $r_{200}$ measured in the 3-D
mass distribution underestimates the amount of mass in the cluster, and mean 
value of the 2-D profiles is a more accurate measurement of the cluster virial 
mass.

The apparent paradox of the best fit NFW profiles to the shear profiles
providing the correct virial masses of the clusters while over-predicting the 
surface densities, from which the shear profiles are calculated, is a result
of the mass sheet degeneracy.  As can be seen in Eq.~\ref{eq5}, $\kappa$ profiles
related by
\begin{equation}
\kappa ^{\prime}(r) = \kappa (r) + (1 - \kappa (r))\times \lambda,
\label{eq9}
\end{equation}
for any constant $\lambda$, produce the same shear profile.  This is
due to the shear profiles measuring the change in mass with radius, and
therefore weak lensing only being able to measure the mass relative to the
density at the outer radius of the measured shear region.  The imposition
of a chosen mass model breaks the mass sheet degeneracy, provided the model
has a bijective relation between the shape of the surface density profile and the total
mass at a given radius.  There is nothing, however, which prevents an 
incorrect model from being assumed, and therefore measuring a mass profile
which differs from the true profile by some value of $\lambda$ via Eq.~\ref{eq9}.
In this case, while the NFW model does
provide a relation between the surface density slope, which is measured by
the shear profile, and the total mass, which is not, at the outer edge
of the shear profile, the mass density assumed at large radii is incorrect, 
and the best fit models over predict the total surface density within the
fitting region.

Our result that the mass measured by weak lensing observations is affected
mostly by the alignment of the line-of-sight to the major-axis of the
cluster, and therefore the small-scale substructure is of minimal importance, 
is in good agreement with the results of
\citet{KI01.1}.  This suggests that for the purposes of modeling weak lensing
observations, clusters can be adequately described by a smooth, tri-axial
mass distribution.  Massive sub-halos projected onto the cluster core can, 
however, perturb the measured concentration parameter for the cluster.  The
levels of the perturbations of the concentration were typically $\sim 10-20\%$,
except in the case of the major merger cluster g72, in which case the perturbations
were on the order of $50-100\%$.  Infalling haloes which are outside the virialized
region of the cluster can also increase the measured $r_{200}$, as occurred for
one projection of g51, but such projections should be detected in redshift
surveys \citep[e.g.][]{CZ02.1}.

\begin{figure}
\centering
\resizebox{\hsize}{!}{\includegraphics{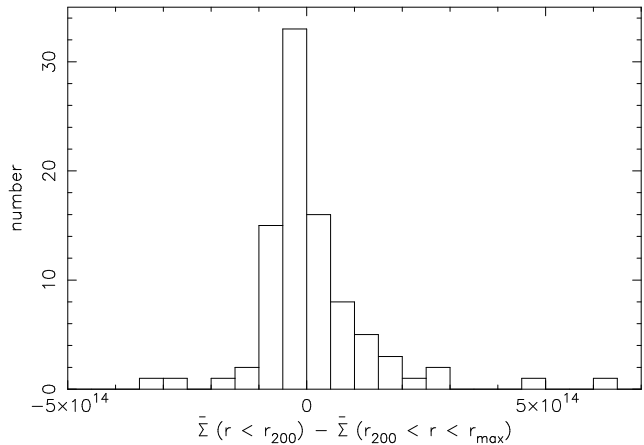}}
\caption{Plotted above are the amounts of excess surface mass within
$r < 3 \mathrm{Mpc}$ compared to an annular region of $3 \mathrm{Mpc} < r <
4 \mathrm{Mpc}$ in projections of a 128 Mpc box with the inner 13 Mpc 
removed.}
\label{fig10}
\end{figure}

Our results that we find, on average, the correct $r_{200}$, and therefore
the correct virial mass, for the clusters is in stark contrast to the results
of \citet{CE97.1} and \citet{ME01.1}, who found that weak lensing measurements
would be consistently higher than the virial masses of the clusters.  The
two major differences between our study and theirs are the technique used to 
measure mass via weak lensing and the size of the projected line-of-sight 
through the clusters.

Both \citet{CE97.1} and \citet{ME01.1} simulated weak lensing observations
by calculating the aperture densitometry statistic
\begin{equation}
\zeta = \bar{\kappa}(r<r_{\mathrm{ap}}) - 
\bar{\kappa}(r_{\mathrm{ap}}<r<r_{\mathrm{max}})
\end{equation}
which is the mean $\kappa$ within some radius $r_{\mathrm{ap}}$ minus the
mean $\kappa$ in an annular region from $r_{\mathrm{ap}}$ to 
$r_{\mathrm{max}}$ \citep{FA94.1}, which can be measured by convolving
the shear profile with a specific kernel.  Rather than calculating the shear
produced by the clusters, both papers estimated $\zeta$ directly from the
projected images by measuring the $\bar{\kappa}$ within a given radius.
However, instead of subtracting off the mean $\kappa$ within an annular
region immediately surrounding the radius used to measure $\bar{\kappa}$,
they subtracted off the mean $\kappa$ measured across all of the simulations.
Given that the surface density is still decreasing with increasing radius
at $r = r_{200}$, the mean $\kappa$ in an annular region immediately outside
the $r = r_{200}$ aperture is higher than the mean surface density in the rest
of the simulated regions.  As a result, both papers overestimated $\zeta$ for
their clusters, and therefore overestimated the cluster mass which would have
been measured by this weak lensing technique.

Both papers then compared the $\zeta$ values, converted into
a surface mass, with cluster models.  With this method, the chosen cluster 
model must have the correct mass profile at all radii, as one must integrate
along the line of sight to measure the surface mass density, in order make 
an accurate comparison.
The 5-10$\%$ overestimate of the cluster mass from this technique in 
\citet{CE97.1} is comparable in size, although opposite in sign, to the 
$\sim 5\%$ difference between surface mass at $r_{200}$ of the the NFW 
profiles used here and the true surface mass of the simulation.  
The larger overestimate of cluster mass via weak lensing from 
\citet{ME01.1} is a result of their choice of a cluster model, namely a
spherical model
which has a mean density of $200\times \rho_{\mathrm{crit}}$ inside a sphere
with radius $r_{200}$, and a zero density outside the sphere.  The excess mass
which is detected is therefore likely to be that associated with the cluster
outside of $r_{200}$.  Indeed, as can be seen in Fig.~\ref{fig9}, if
\citet{ME01.1} had integrated an NFW profile to the edge of their simulated
regions instead of only out to $r_{200}$ to measure the expected surface mass
of the clusters, they would have found that the technique was underestimating,
instead of overestimating, the cluster mass.

While we project the cluster in a 13 Mpc box, the projected regions of
\citet{CE97.1} (64 Mpc box) and \citet{ME01.1} (128 Mpc sphere) are much 
larger.  If there is a structure along the line of sight, such as a filament, 
then the surface mass of the field will increase, and such structures might be
more common in the regions around massive clusters.  Indeed, this is one of 
the reasons
quoted in \citet{ME01.1} for why they find a greater over-estimate of the
mass via weak lensing than was found in \citet{CE97.1}.

In order to test if the 13 Mpc box size causes us to underestimate the mass
which would be measured with our weak lensing technique, we selected three 
clusters from the original simulation the
re-simulations were based on.  For each cluster, we rotated the particles
to obtain 10 independent lines-of-sight, cut out a 128 Mpc box around the
cluster, and projected along the three sides to give 30 independent projections
per cluster.  We then cut out a 13 Mpc box around each cluster, performed
the same projections, and subtracted them from the 128 Mpc box projections
to obtain the projection of only those structures which would be located
outside of the high-resolution region used in the NFW fits.
For each projection we then calculated the $\zeta$ statistic
with $r_{\mathrm{ap}} = 3$ Mpc, the outer radius of our shear fitting region,
and $r_{\mathrm{max}} = 4$ Mpc.  

These values of $\zeta$, converted
into a surface mass by multiplying by $\pi r_{\mathrm{ap}}^2 \Sigma 
_{\mathrm{crit}}$, are shown as a histogram in Fig.~\ref{fig10}.  The mean of the 
$\zeta$ distribution is consistent with zero, although 
with a skew that results in a broader tail to the large positive surface
masses.  The distribution has a rms of $1.2\times 10^{14} \mathrm{M}_\odot$,
which is $\sim 5\%$ of the surface mass at this radius for the simulated
clusters.  The two projections with high values for the increased mass within
$r_{\mathrm{ap}}$ are both the result of a second, smaller cluster along the 
line-of-sight. 

Thus, expanding the projected region from our original 13 Mpc to
128 Mpc would have only resulted in a small scatter being added to the
measured cluster masses, with the exception of the occasional projection
which would have a significantly higher mass (up to $25\%$ additional mass,
although these simulations were chosen to not have similar mass neighbors so 
exclude the possibility of two equal mass clusters being projected along
the line of sight).  These projections with
higher projected mass, however, are caused primarily by secondary
clusters along the line-of-sight, and should be visible in the form of a 
concentration of galaxies at redshifts slightly different from the main
cluster. 

\section{Summary}
We have fit NFW profiles to 30 surface density projections for each of 4 
simulated clusters, and have found that the line-of-sight variations in the
projections can lead to dispersions in the parameters of the best-fit models
on the order of the errors expected in high-quality weak lensing
observations.  Most of the dispersion is due to the tri-axial 
nature of the clusters, and how close the line-of-sight for the projection was
to the major-axis of the cluster.  For all of the projections, a NFW surface
density profile provided a good enough fit so that, with the expected errors
in a high-quality weak lensing observation, an observer would measure a
reduced $\chi ^2$ close to 1.  

Further, there is a general correlation 
between the two parameters in the NFW models, $r_{200}$ and $c$.  This
correlation is due largely to the direction of the major-axis of the clusters 
not varying largely with radius, and thus a line-of-sight near the major
axis would have both a large amount of mass projected onto the core of the
cluster and an overall increase in the surface density of the cluster at
all radii within $r_{200}$.  Additionally, the projection of sub-halos 
outside the core of the cluster onto the core causes an additional scatter
in the best-fit values for the concentration.  The level of scatter in
the $r_{200}$ and $c$ best-fit models is comparable to the error expected
in weak lensing observations using current telescopes in ideal conditions.

There is also an anti-correlation detected between the best-fit $r_{200}$
for a projection and the ellipticity of the cluster in the projection.  Because
the of the variation in ellipticities of individual clusters, however, no correction
for the measured $r_{200}$ based on the measured ellipticity is possible.

The shear fields from the 30 projections for each cluster were averaged, and
the best-fit NFW profile was measured.  For three of the four clusters, the
difference in the value for $r_{200}$ between the 2-D averaged fit and the 3-D
fit is within the errors expected due to the finite number of projections.
We argue that the difference in the discrepant cluster is a result of the
3-D spherical fit underestimating the virial mass of the cluster due to
the high ellipticity of the cluster.

We found that while the NFW profile fitting technique correctly
measured the virial mass on average, the predicted surface mass in the images
were all overestimated by the best-fit parameters.  This is due to the mass
density at large radius decreasing faster than the $r^{-3}$ assumed by the 
NFW model.  While the NFW model was defined only out to $r_{200}$, calculations
of the surface mass observed with weak lensing requires integration along the
line of the sight of all the mass in the field, even if it has not yet fallen
into the cluster.  The overestimate of the 3-D density of the NFW profile when
integrated beyond $r_{200}$ results in a significant overestimate of the surface
density of the clusters at radii as small as one-half of $r_{200}$.
The mass sheet degeneracy, however, allows the model to
overestimate the surface density while still providing a good fit to the
shear profile.  

Finally, we caution that while we expect these results to be true qualitatively for
clusters of all masses and redshift ranges, the quantitative results in the
figures should be used only as estimates of the magnitude of the effects for clusters
of similar mass and redshift as the simulations.  

\section*{Acknowledgments}
Barbara Lanzoni, Felix Stoehr, Bepi Tormen and Naoki Yoshida
are warmly thanked for all the effort put in the re--simulation
project and for letting us use their simulations.
We also wish to thank Peter Schneider,
Felix Stoehr, and Simon White for useful discussions.  This work was supported
by the Deutsche Forschungsgemeinschaft under the project SCHN 342/3--1
(D.~C.~and L.~K.), the Alexander von Humboldt Foundation, the German Federal
Ministry of Education and Research, and the Program for Investment in the
Future (ZIP) of the German Government (G.~D.~L.).

\bibliographystyle{mn2e}
\bibliography{nbody}

\label{lastpage}

\end{document}